# Robust Reactive Power Optimization and Voltage Control Method for Active Distribution Networks via Dual Time-scale Coordination

Weiye Zheng, Wenchuan Wu, Boming Zhang, Yongjie Wang
(Department of Electrical Engineering, Tsinghua University, China )

*Abstract*—In distribution networks, there are slow controlling devices and fast controlling devices for Volt-VAR regulation. These slow controlling devices, such as capacitors or voltage regulators, cannot be operated frequently and should be scheduled tens of minutes ahead (Hereafter named as slow control). Since of the uncertainties in predicting the load and distributed generation, the voltage violations cannot be eliminated by fast controlling devices with improper schedule of the slow controlling devices. In this paper we propose dual time-scale coordination for the Volt-VAR control scheme, corresponding to slow and fast control. In the case of slow control, a robust voltage and reactive power optimization model is developed. This guarantees that subsequent fast controls can maintain the system's voltage security if the uncertain parameters vary within predefined limits. This nonconvex optimization problem is relaxed to a mix integer second order conic problem, and the dual form of its sub-problem is also derived. Then a column-and-constraint generation algorithm was used to solve the robust convexified model. A conventional deterministic optimization model can be used to determine the fast control mechanism. Numerical tests were conducted on a real distribution feeder in China, a balanced IEEE 69-bus and unbalanced 123-bus benchmark distribution networks. The simulation results show that solving the deterministic model is not always feasible and voltage violation may occur. The robust model was shown to be effective with respect to all possible scenarios in the uncertainty set, with little compromise in terms of network losses.

## NOMENCLATURE

The main symbols used throughout this paper are stated below for quick reference. Others are defined as required.

*A. Indices and Sets*

| | |
|---|---|
| $i,j$ | Indices of buses, from 1 to $N$. |
| $\phi$ | Indices of phases, typically referring to *a*, *b* and *c*. |
| $t$ | Iteration of the proposed algorithm. |
| $k$ | Index of the tap settings of voltage regulators, from 1 to $R$. |
| $\mathcal{N}$ | Set of buses in the entire system. |
| $\mathcal{E}$ | Set of branches in the entire system. |
| $\mathcal{K}_{ij}$ | Set of allowable taps of branch *ij*, consisting of $\{K_{ij,1}, K_{ij,2}, ..., K_{ij,R}\}$.. |
| $\mathcal{B}_i$ | Set of allowable on-off decisions for capacitor banks at bus *i*. |
| $\mathcal{X}$ | Feasible region of the first stage variables. |
| $\mathcal{VR}$ | Set of branches with voltage regulators. |
| $\mathcal{MSC}$ | Set of buses with machanically switched capacitors/capacitor banks. |
| $\mathcal{D}$ | Uncertainty set for fluctuation of ***d***. |

*B. Input Parameters and Functions*

| | |
|---|---|
| $r_{ij}, x_{ij}$ | Resistance and reactance of branch *ij*, respectively. |
| $V_{ref}$ | Voltage at the point of common coupling |
| $\bar{P}_{ij}, \underline{P}_{ij}$ | Upper and lower bounds of the active power at the sending end of branch *ij*, respectively. |
| $\bar{Q}_{ij}, \underline{Q}_{ij}$ | Upper and lower bounds of the reactive power at the sending end of branch *ij*, respectively. |
| $\bar{I}_{ij}$ | Upper bound of the current magnitude of branch *ij*. |
| $\overline{V}_i, \underline{V}_i$ | Upper and lower bounds of the voltage magnitude at bus *i*, respectively. |
| $\overline{Q}_{Ci}, \underline{Q}_{Ci}$ | Upper and lower bounds of the injected reactive powers of the static var compensation devices at bus *i*, respectively. |
| $\overline{Q}_{Gi}, \underline{Q}_{Gi}$ | Upper and lower bounds of the injected reactive powers of the generators at bus *i*, respectively. |
| $\overline{b}_i^C$ | The admittance element of capacitor bank at bus *i*. |
| $d_0$ | Forecasted loads and active outputs of DGs |
| $d$ | Realized loads and active outputs of DGs |

## C. Variables

| | |
|---|---|
| $P_{ij}, Q_{ij}, l_{ij}$ | Active power flow, reactive power flow and squared magnitude of the current at the sending end of branch *ij*, respectively. |
| $P_{Di}, Q_{Di}$ | Active and reactive power load demands at bus *i* respectively. |
| $P_{Gi}, Q_{Gi}$ | Injected active and reactive powers of generators at bus *i*. |
| $Q_{Ci}$ | Total reactive power from compensators at bus *i*. |
| $v_i$ | Squared voltage magnitude at bus *i*. |
| $P_i, Q_i$ | Total injected active and reactive powers at bus *i*. |
| $\kappa_{ij}$ | Turn ratio variable for tap changer at branch *ij*. |
| $\beta_i$ | On and off decision variable for the capacitor bank at bus *i*. |
| $\boldsymbol{\beta}$ | On and off decision vector for capacitor banks. |
| $\boldsymbol{\kappa}$ | Tap setting vector for voltage regulators. |
| $\boldsymbol{x}$ | First-stage decision vector for voltage regulators and capacitor banks. |
| $\boldsymbol{y}$ | Second-stage decision vector for reactive outputs of static VAR compensators and distributed generators. |

*Notation:* Upper (lower) boldface letters will be used for matrices (column vectors); $\|\cdot\|_p$ denotes the vector *p*-norm for $p \geq 1$; $(\cdot)^T$ transposition; $\boldsymbol{I}$ the identity matrix; variables with superscript $\varphi$ denote corresponding variables for phase $\varphi$.

## I. INTRODUCTION

Volt-VAR optimization aims to minimize network power losses and prevent voltage violations by dispatching reactive power control devices [1]. In recent years, the numbers of renewable distributed generations (DGs), including photovoltaic (PV) and wind power generators, have increased significantly in distribution networks, formulating active distribution networks (ADNs). Due to the reverse power flow, DG outputs may lead to overvoltage [2]. To address this issue, reactive power optimization was investigated in [3], in which it was shown that power losses can be reduced significantly. Most studies concerning reactive power optimization do not consider any uncertain factors [4]-[6].

Three major factors contribute to the uncertainties in ADNs:

1) The active outputs of renewable DGs fluctuate dramatically due to their inherent volatility and intermittency;

2) Load demands vary with time and are hard to predict precisely;

3) There are limited real-time measurements in distribution networks, so large errors may occur in the estimations of the outputs of the DGs and load demands.

These uncertainties pose a series of technical challenges to the operation of ADNs. Voltage regulation in distribution networks is conventionally accomplished by voltage regulators (VRs), under-load tap changing (ULTCs) transformers and capacitor banks/mechanically switched capcaitors (MSCs). Although these devices are effective in managing slow variations in voltages, on the time scale of hours, they do not perform well for fast fluctuations, on the time scale of minutes or seconds.

On the one hand, these slow controlling devices cannot be operated frequently and they are expected to maintain their states for a long time, since the maximum allowable daily operating time (MADPT) is limited by the expected lifetime. On the other hand, in certain scenarios, where the actual loads and outputs of the DGs significantly deviate from their predicted values, voltage violation may occur. Some recent papers have made efforts to address these uncertainties. Based on Niederreiter's quasi-random sampling technique, the authors of [7] proposed a robust algorithm for Volt-VAR control. However, optimality is not guaranteed by the theory. A robust voltage control model is described in [8], where a scenario generation and reduction method is used, but slow controlling devices such as ULTCs and MSCs are not considered. Based on scenario generation and reduction method, [9] proposed a robust voltage regulation method, where monotonicity assumption is necessary for scenario reduction. Also, the discrete variables introduced by ULTCs and MSCs are tackled with an empirical two-step solution method.

This paper tackles the uncertainties by coordinating slow and fast controlling devices. A dual time scale coordinated control scheme is proposed, as shown in Figure 1.

In the case of the larger time scale, here-and-now dispatch of slow controlling devices (including VRs and MSCs) is determined. This occurs on scales of tens of minutes or even hours (hereafter referred to as **slow control**). In the case of the smaller time scale, the controlling device is activated every minute using a wait-and-see control method for reactive optimal power flow, e.g., reactive outputs of DGs and SVCs, are used to mitigate the impact of frequent voltage variations that occur due to uncertain demands and the output of the DGs (hereafter referred to as **fast control**). In the case of slow control, a robust Volt-VAR optimization model is proposed to determine the operation of slow devices and guarantee that the subsequent fast controls can ensure the system's voltage security as long as the uncertain parameters vary within the predefined limits. As conventional deterministic optimization models are suitable for fast control, this paper addresses a robust optimization model used for slow control.

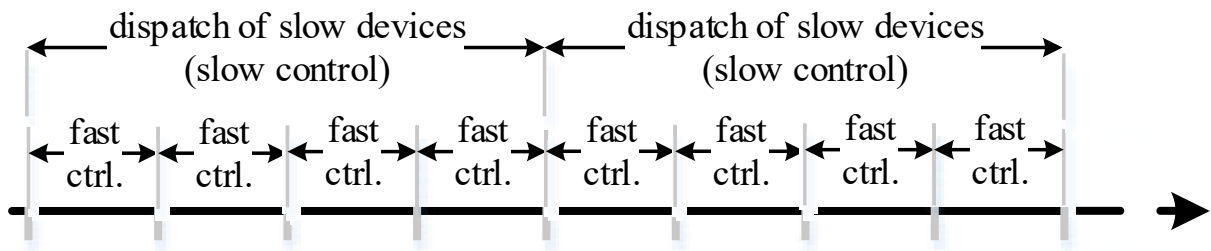

Figure 1 Timeline for sequential coordination of slow control and fast control

As shown in Figure 1, the operating times for the slow controls should be optimally scheduled in advance. Some heuristic approaches have been proposed in previous work [10][11]. This paper concentrates on how to generate optimal control strategies when the operating times are determined. Since the MADPTs of the slow controlling devices are considered when scheduling operating times, the sequential slow controls can be formulated as independent optimization problems.

The main contributions of this paper include:

(1) No literatures before have ever reported such a dual time scale coordinated control scheme to reduce the uncertainty of load and distributed generation for three phase distribution networks. Different from [9] which is essentially based on scenario method, the coordination is directly cast as a multilevel robust optimization problem in this paper. Both the convergence of the dual time-scale coordination and the optimality under uncertainties are guaranteed theoretically.

(2) A robust Volt-VAR optimization model is developed to determine the operation of slow devices and guarantee that the subsequent fast controls can maintain the system's voltage security under specified uncertainties.

The remainder of the paper is organized as follows: In Section II, the robust Volt-VAR optimization model for slow control is formulated. Section III describes the solution methodology. Section IV details the results of numerical simulations. These demonstrate the performance of the proposed algorithm. Section V concludes the paper.

## II. ROBUST VOLT-VAR OPTIMIZATION MODEL

In this section, we describe deterministic and robust Volt-VAR optimization models. First, we review a deterministic model using forecast loads and active DG outputs.

### A. Deterministic Model

For an ADN with $N$ buses, where bus 1 denotes the point of common coupling, the reactive power optimization problem is to find an optimal dispatch for the system equipment (e.g., VRs, MSCs, DGs and SVCs, etc.). This should minimize losses whilst taking into account security constraints. Farivar *et al.* [12] showed that the non-convex branch flow in radial distribution networks can be convexified and that the deterministic model can be formulated as the following problem.

*1) Objective function*

We aim to minimize the active power loss of the network, i.e.,

$$\min_{\beta, \kappa, Q_C, Q_G} \sum_{\varphi=a,b,c} \sum_{(i,j)\in\mathcal{E}} l_{ij}^\varphi r_{ij}^\varphi . \quad (1)$$

*2) DistFlow branch flow constraints in second-order conic programming (SOCP) form*

In power distribution systems, DistFlow branch equations are typically used to describe the power flow. The equations were first proposed by Baran and Wu in [13] and later adopted by Steven H. Low in [12] for relaxations and convexification.

$$\sum_{i:i\to j}(P_{ij}^\varphi - l_{ij}^\varphi r_{ij}^\varphi) + P_j^\varphi = \sum_{k:j\to k} P_{jk}^\varphi, \ \forall j\in\mathcal{N}, \forall \varphi \quad (2)$$

$$\sum_{i:i\to j}(Q_{ij}^\varphi - l_{ij}^\varphi x_{ij}^\varphi) + Q_j^\varphi - \beta_j^\varphi \overline{b}_j^{\varphi,C} v_j^\varphi = \sum_{k:j\to k} Q_{jk}^\varphi, \ \forall j\in\mathcal{N}, \forall \varphi \quad (3)$$

$$v_j^\varphi = \kappa_{ij}^{\varphi,2} v_i^\varphi - 2(r_{ij}^\varphi P_{ij}^\varphi + x_{ij}^\varphi Q_{ij}^\varphi) + (r_{ij}^{\varphi,2} + x_{ij}^{\varphi,2}) l_{ij}^\varphi, \ \forall (i,j)\in\mathcal{E}, \forall \varphi$$

$$(4)$$

$$\left\| \begin{array}{c} 2P_{ij}^\varphi \\ 2Q_{ij}^\varphi \\ l_{ij}^\varphi - v_i^\varphi \end{array} \right\|_2 \leq l_{ij}^\varphi + v_i^\varphi, \ \forall (i,j)\in\mathcal{E}, \forall \varphi \quad (5)$$

$$P_j^\varphi = P_{Gj}^\varphi - P_{Dj}^\varphi, \ \forall j\in\mathcal{N}, \forall \varphi \quad (6)$$

$$Q_j^\varphi = Q_{Gj}^\varphi + Q_{Cj}^\varphi - Q_{Dj}^\varphi, \ \forall j \in \mathcal{N}, \forall \varphi \quad (7)$$

$$\kappa_{ij}^\varphi \in \mathcal{K}_{ij} \quad (8)$$

$$\mathcal{K}_{ij} = \begin{cases} \{K_{ij,1}, K_{ij,2}, ..., K_{ij,R}\}, & \text{if } (i,j) \in \mathcal{VR} \\ \{1.0\}, & \text{otherwise} \end{cases} \quad (9)$$

$$\beta_j^\varphi \in \mathcal{B}_j \quad (10)$$

$$\mathcal{B}_j = \begin{cases} \{0,1\}, & \text{if } j \in \mathcal{CB} \\ 0, & \text{otherwise} \end{cases}. \quad (11)$$

Here, (2) is the conventional nodal active power balance constraint used in the DistFlow equations. (3) extends the conventional nodal reactive power balance constraint by considering MSCs. (4) is a general form of the voltage constraint that takes into account branches with VRs, as shown in Figure 2. (5) is the relaxed second-order conic constraint proposed in [12]. (6) and (7) represent the net injected active and reactive power for each bus respectively. (8) and (9) describe all allowable taps of branch $ij$; (10) and (11) specify allowable on-off decisions at bus $i$. Combined with equations (8)-(11), the branch flow equations in (3) and (4) provide a unified expression for the cases with or without MSCs and VRs.

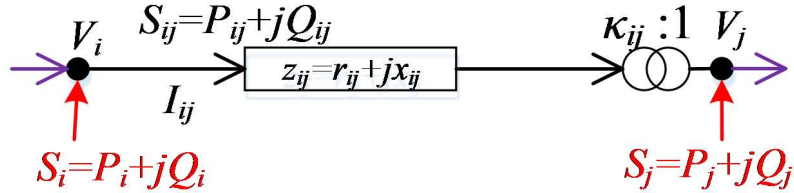

Figure 2  DistFlow branch flow model with voltage regulators

3) *Reference voltage of point of common coupling (PCC)*

$$V_1^\varphi = V_{ref}, \forall \varphi. \quad (12)$$

(12) implies that the voltage of PCC is set to reference voltage.

4) *Security constraints*

$$\left|l_{ij}^\varphi\right| \le \overline{I}_{ij}^{\varphi,2}, \ \forall (i,j) \in \mathcal{E}, \forall \varphi \quad (13)$$

$$\underline{P}_{ij}^\varphi \le P_{ij}^\varphi \le \overline{P}_{ij}^\varphi, \ \forall (i,j) \in \mathcal{E}, \forall \varphi \quad (14)$$

$$\underline{Q}_{ij}^\varphi \le Q_{ij}^\varphi \le \overline{Q}_{ij}^\varphi, \ \forall (i,j) \in \mathcal{E}, \forall \varphi \quad (15)$$

$$\underline{V}_i^{\varphi,2} \le v_i^\varphi \le \overline{V}_i^{\varphi,2}, \ \forall i \in \mathcal{N}, \forall \varphi. \quad (16)$$

The security constraints for the current, active power flow and reactive power flow of branch $ij$, are represented by (13)-(15), respectively. Voltage violation is prevented by introducing a voltage constraint (16).

5) Operating constraints of DGs and compensation devices

$$\underline{Q}_{Gi}^{\varphi} \leq Q_{Gi}^{\varphi} \leq \overline{Q}_{Gi}^{\varphi}, \ \forall i \in \mathcal{N}, \forall \varphi \quad (17)$$

$$\underline{Q}_{Ci}^{\varphi} \leq Q_{Ci}^{\varphi} \leq \overline{Q}_{Ci}^{\varphi}, \ \forall i \in \mathcal{N}, \forall \varphi. \quad (18)$$

The reactive powers of the DGs and SVCs are confined by (17)-(18).

B. Convexification of deterministic model

It is very difficult to obtain global optimal solutions of the nonconvex deterministic model efficiently. Therefore, the cross terms in (3) and (4) have to be relaxed.

1) Relaxation of MSCs

By introducing $\omega_j^{\varphi} = v_j^{\varphi} \beta_j^{\varphi}$, [4] shows that (3) is identical to

$$\sum_{i:i \to j}(Q_{ij}^{\varphi} - l_{ij}^{\varphi} x_{ij}^{\varphi}) + Q_j^{\varphi} - \overline{b}_j^{\varphi,C} \omega_j^{\varphi} = \sum_{k:j \to k} Q_{jk}^{\varphi}, \ \forall j \in \mathcal{N}, \forall \varphi \quad (19)$$

with additional constraints:

$$v_j^{\varphi} - \overline{V}_j^{\varphi,2}(1-\beta_j^{\varphi}) \leq \omega_j^{\varphi} \leq v_j^{\varphi} - \underline{V}_j^{\varphi,2}(1-\beta_j^{\varphi}), \ \forall j \in \mathcal{N}, \forall \varphi$$

$$(20)$$

$$\underline{V}_j^{\varphi,2} \beta_j^{\varphi} \leq \omega_j^{\varphi} \leq \overline{V}_j^{\varphi,2} \beta_j^{\varphi}, \ \forall j \in \mathcal{N}, \forall \varphi. \quad (21)$$

2) Exact linearization of VR model

The intuitive idea is to approximate the original surface by a series of polyhedra. The approximation is exact in the case of discrete feasible set of tap settings for VRs, as shown in Figure 3. By using the exact piecewise linearization technique from [14], (4) can be rewritten as

$$\sum_{k=1}^{R-1} K_{ij,k}^{\varphi,2}(g_{ij,k}^{\varphi} \underline{V}_i^{\varphi,2} + h_{ij,k}^{\varphi} \overline{V}_i^{\varphi,2}) - 2(r_{ij}^{\varphi} P_{ij}^{\varphi} + x_{ij}^{\varphi} Q_{ij}^{\varphi})$$
$$+(r_{ij}^{\varphi,2} + x_{ij}^{\varphi,2}) l_{ij}^{\varphi} - v_j^{\varphi} = 0, \ \forall (i,j) \in \mathcal{E}, \forall \varphi \quad (22)$$

with extra constraints:

$$\sum_{k=1}^{R}(g_{ij,k}^{\varphi} + h_{ij,k}^{\varphi}) = 1 \quad (23)$$

$$\sum_{k=1}^{R-1} \delta_{ik}^{\varphi} = 1 \quad (24)$$

$$0 \leq g_{ij,1}^{\varphi}, h_{ij,1}^{\varphi} \leq \delta_{ij,1}^{\varphi} \quad (25)$$

$$0 \leq g_{ij,R}^{\varphi}, h_{ij,R}^{\varphi} \leq \delta_{ij,R-1}^{\varphi} \quad (26)$$

$$0 \le g_{ij,k}^{\varphi}, h_{ij,k}^{\varphi} \le \delta_{ij,k}^{\varphi} + \delta_{ij,k-1}^{\varphi}, \forall k \in [2, R-1] \quad (27)$$

$$\delta_{ij,k}^{\varphi} \in \{0,1\}. \quad (28)$$

where $\delta_{ik}$ denotes the section of interest. Note that the allowable taps divide the function $f$ into $R$ sections, $g_{ij,k}$ and $h_{ij,k}$ are auxiliary variables that approximate the original function $f$. For details, please refer to [14].

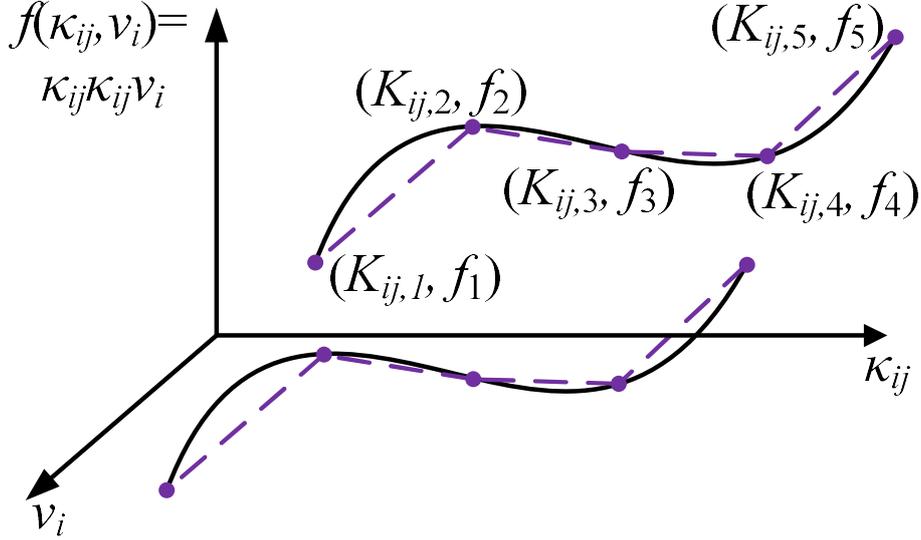

Figure 3 Piecewise linearization of cross term introduced by Voltage Regulators (VRs)

3) *Compact form of the convex deterministic Volt- VAR Optimization model (DVO)*

The convex deterministic model consists of an objective

$$\min_{\beta,\kappa,g,h,\delta,\omega,Q_C,Q_G} \sum_{\varphi=a,b,c} \sum_{(i,j)\in\mathcal{E}} l_{ij}^{\varphi} r_{ij}^{\varphi} \quad (29)$$

with constraints (2), (19)-(21), (22)-(28) and (5)-(18). By defining an augmented vector $x = (\beta,\kappa,g,h,\delta,\omega)$ that holds the decision variables for the MSCs and VRs, a vector $y = (Q_C, Q_G, v, l_{ij}, P_{ij}, Q_{ij})^T$ that relates the power flow and the dispatch of DGs and SVCs and a forecasted vector $d_0 = (P_D, Q_D, Q_G)^T$, this model can also be written in a compact form:

$$\min_{x,y} \quad b^T y \quad (30)$$

s.t.
$$Ax + By + d_0 = 0 \quad (31)$$

$$Cx + Dy \ge e \quad (32)$$

$$x \in \mathcal{X} \quad (33)$$

$$\|G_l y\|_2 \le g_l^T y, \forall l \in \mathcal{E} \quad (34)$$

$$\underline{y} \le y \le \overline{y}. \quad (35)$$

where (30) is equivalent to (29), (31) represents the power balance constraints (2), (19) and (22), (32) is identical to the inequality constraints (20)-(21), (33) is an expression for the constraints (8)-(11) and (23)-(28), (34) simplifies the expression of (5), and the security constraints in (13)-(16) and the reference constraint in (12) can be represented by (35).

*C. Two-Stage Robust Volt-VAR Optimization Model (RVO)*

In the deterministic model, neither the difference in time-scales among our devices nor the uncertainties in the loads and DG outputs is considered. In the robust model, it is assumed that the dispatch for the slow control (VRs and MSCs) is decided prior to the loads and active DG outputs of DGs occurring. For each instance of uncertain loads and DG outputs, the fast control (reactive outputs of SVCs and DGs) reduces network losses and prevents voltage violation. The first-stage dispatch of VRs and MSCs attempts to minimize the worst-case second-stage power loss, which can be formulated as the following two-stage robust model:

$$\min_{x} \max_{d \in \mathcal{D}} \min_{y} \quad \boldsymbol{b}^T \boldsymbol{y} \quad (36)$$

$$\text{s.t.} \quad \mathbf{A}\boldsymbol{x} + \mathbf{B}\boldsymbol{y} + \boldsymbol{d} = 0, \quad \forall \boldsymbol{d} \in \mathcal{D} \quad (37)$$

$$\mathbf{C}\boldsymbol{x} + \mathbf{D}\boldsymbol{y} \geq \boldsymbol{e}, \quad \forall \boldsymbol{d} \in \mathcal{D} \quad (38)$$

$$\|\mathbf{G}_l \boldsymbol{y}\|_2 \geq \boldsymbol{g}_l^T \boldsymbol{y}, \quad \forall l \in \mathcal{E}, \forall \boldsymbol{d} \in \mathcal{D} \quad (39)$$

$$\underline{\boldsymbol{y}} \leq \boldsymbol{y} \leq \overline{\boldsymbol{y}}, \quad \forall \boldsymbol{d} \in \mathcal{D} \quad (40)$$

and (33).

The uncertainty set $\mathcal{D}$ is assumed to form a polyhedron. This guarantees that the column-and-constraint generation (C&CG) algorithm used to solve the proposed model converges after a finite number of iterations [15]. First we calculate the optimal reactive power flow of a given schedule $\tilde{x}$ of VRs and MSCs for a particular load and DG scenario $\boldsymbol{d}$, which replaces the forecasted value $\boldsymbol{d}_0$. This enables us to obtain the actual power losses. Next, the worst scenario where the optimal losses are maximized for a given dispatch of VRs and MSCs is obtained. Finally, our robust model finds an optimal way to dispatch the VRs and MSCs. This minimizes the worst-case losses. It is ensured by (37)-(40) that power balance and security constraints are maintained when loads and active outputs of DGs fluctuate within the uncertainty set $\mathcal{D}$.

III. SOLUTION METHODOLOGY

The C&CG algorithm in [15] is used to solve the two-stage robust reactive power optimization model. The distinctive feature of the proposed robust model is that the innermost problem is an SOCP, while the outermost problem is a mixed integer second-order conic program (MISOCP). The C&CG algorithm can be naturally extended for this robust reactive power optimization model. The application of the algorithm is outlined as follows. The C&CG algorithm tries to solve the following extensive formulation of the proposed robust model:

$$\min_{x,y(d),\eta} \eta \qquad (41)$$

$$\text{s.t.} \qquad b^T y(d) \leq \eta \qquad (42)$$

$$\eta \geq 0 \qquad (43)$$

$$Ax + By(d) + d = 0, \forall d \in \mathcal{D} \qquad (44)$$

$$Cx + Dy(d) \geq e, \quad \forall d \in \mathcal{D} \qquad (45)$$

$$\|G_l y(d)\|_2 \geq g_l^T y(d), \quad \forall l \in \mathcal{E}, \forall d \in \mathcal{D} \qquad (46)$$

$$\underline{y} \leq y(d) \leq \overline{y}, \quad \forall d \in \mathcal{D} \qquad (47)$$

and (33).

The above problem has an infinite number of variables and constraints corresponding to each scenario $d$.

*A. Master Problem*

Due to the infinite number of constraints (44)-(47), which are indexed by the uncertainty set $\mathcal{D}$, it is not possible to solve the above problem directly. The following reduced problem, where $\mathcal{D}$ is replaced by a finite subset, provides a lower bound for the robust model:

$$\min_{x,y(t),\eta} \eta \qquad (48)$$

$$\text{s.t.} \qquad by(t) \leq \eta, \quad \forall t = 1, 2, ..., t_0 \qquad (49)$$

$$Ax + By(t) + d(t) = 0, \forall t = 1, 2, ..., t_0 \qquad (50)$$

$$Cx + Dy(t) \geq e, \quad \forall t = 1, 2, ..., t_0 \qquad (51)$$

$$\|G_l y(t)\|_2 \geq g_l^T y(t), \forall l, \forall t = 1, 2, ..., t_0 \qquad (52)$$

$$\underline{y} \leq y(t) \leq \overline{y}, \quad \forall t = 1, 2, ..., t_0 \qquad (53)$$

and (33).

where $d(t) \in \mathcal{D}$, $t = 1, 2, ..., t_0$. The master problem is an MISOCP with binary variables in $x$ and second-order conic constraints in (52).

*B. Sub-problem*

The next key step in the C&CG algorithm is to search for the worst-case scenario for the master problem and obtain an upper bound for the robust model. This is achieved by solving the following sub-problem for the given dispatch scheme $x^*$ of VRs and MSCs:

$$\eta(x^*) = \max_{d \in \mathcal{D}} \min_{y} b^T y \qquad (54)$$

$$\text{s.t.} \qquad By + d = -Ax^*, \quad (\tau) \qquad (55)$$

$$\mathbf{D}y \geq e - \mathbf{C}x^*, \quad (\xi) \qquad (56)$$

$$\|\mathbf{G}_l y(t)\|_2 \geq g_l^T y(t), \forall l \ (\sigma_l, \mu_l) \qquad (57)$$

$$\underline{y} \leq y \leq \overline{y}, \quad (\underline{\lambda}, \overline{\lambda}) \qquad (58)$$

The optimal objective value of this sub-problem is the worst-case network losses for a given dispatch scheme $x^*$. Hence, it can be used to estimate an upper bound for the robust model. This two-level max-min problem cannot be solved directly. Therefore, we convert the inner conic problem. The Lagrangian function of the inner problem is:

$$\begin{aligned} L(y, o_l, \omega_l, \sigma_l, \underline{\lambda}, \overline{\lambda}, \xi, \mu_l, v_l) &= b^T y - \tau^T (\mathbf{A}x^* + \mathbf{B}y + d) - \sum_l \sigma_l^T (\mathbf{G}_l y - o_l) \\ &\quad - \sum_l \mu_l (g_l^T y - \omega_l) - \sum_l v_l (\omega_l - \|o_l\|_2) - \underline{\lambda}^T (y - \underline{y}) \\ &\quad - \overline{\lambda}^T (\overline{y} - y) - \xi^T (\mathbf{D}y + \mathbf{C}x^* - e) \\ &= [b^T - \tau^T \mathbf{B} - \sum_l (\sigma_l^T \mathbf{G}_l + \mu_l g_l^T) - \underline{\lambda}^T + \overline{\lambda}^T - \xi^T \mathbf{D}]y \\ &\quad - \tau^T (\mathbf{A}x^* + d) + \underline{\lambda}^T \underline{y} - \overline{\lambda}^T \overline{y} + \sum_l (\sigma_l^T o_l + v_l \|o_l\|_2) \\ &\quad + \sum_l (\mu_l - v_l) \omega_l - \xi^T (\mathbf{C}x^* - e) \end{aligned} \qquad (59)$$

with $v_l, \underline{\lambda}, \overline{\lambda}, \xi \geq 0$.

The dual problem can be formulated as:

$$\max_{d, \sigma_l, \underline{\lambda}, \overline{\lambda}, \xi, \mu_l} \inf_{y, o_l, \omega_l} L(y, d, \sigma_l, \underline{\lambda}, \overline{\lambda}, \xi, \mu_l). \qquad (60)$$

The minimization over $y$ and $\omega_l$ is bounded if and only if

$$\mathbf{B}^T \cdot \tau + \sum_l (\mathbf{G}_l^T \sigma_l + \mu_l g_l) + \underline{\lambda} - \overline{\lambda} + \mathbf{D}^T \cdot \xi = b. \qquad (61)$$

$$\mu_l = v_l, \ \forall l \qquad (62)$$

$$\|\sigma_l\|_2 \leq \mu_l, \ \forall l. \qquad (63)$$

To minimize over $o_l$, note that

$$\inf_{o_l} \sigma_l^T o_l + v_l \|o_l\|_2 = \begin{cases} 0, & \|\sigma_l\|_2 \leq v_l \\ -\infty, & \text{otherwise} \end{cases}. \qquad (64)$$

The Lagrangian dual function is

$$\theta(\sigma_l, \underline{\lambda}, \overline{\lambda}, \xi, \mu_l, v_l) = \begin{cases} -(\mathbf{A}x^* + d)^T \tau + \underline{y}^T \underline{\lambda} & \text{if } \|\sigma_l\|_2 \leq v_l, \\ \quad -\overline{y}^T \overline{\lambda} + (e - \mathbf{C}x^*)^T \xi, & (61) \text{ and } (62) \\ -\infty, & \text{otherwise} \end{cases}. \qquad (65)$$

which leads to the dual problem in a monolithic form:

$$\max_{d \in \mathcal{D}, \tau, \underline{\lambda}, \overline{\lambda}, \xi, \sigma_l, \mu_l} -(\mathbf{A}x^* + d)^T \tau + \underline{y}^T \underline{\lambda} - \overline{y}^T \overline{\lambda} + (e - \mathbf{C}x^*)^T \xi \qquad (66)$$

$$\text{s.t.} \quad \mathbf{B}^T \cdot \boldsymbol{\tau} + \sum_l (\mathbf{G}_l^T \boldsymbol{\sigma}_l + \mu_l \mathbf{g}_l) + \underline{\boldsymbol{\lambda}} - \overline{\boldsymbol{\lambda}} + \mathbf{D}^T \cdot \boldsymbol{\xi} = \mathbf{b} \quad (67)$$

$$\|\boldsymbol{\sigma}_l\|_2 \le \mu_l, \quad \forall l \quad (68)$$

$$\boldsymbol{\xi}, \underline{\boldsymbol{\lambda}}, \overline{\boldsymbol{\lambda}}, \mu_l \ge 0 \quad (69)$$

The objective function is linear except for the bilinear term $\mathbf{d}^T \boldsymbol{\tau}$, and the constraints are all in SOCP form. With the assumption that the uncertainty set $\mathcal{D}$ is polyhedral, the extreme points of $\mathcal{D}$ can be described by a set of auxiliary binary variables and linear constraints. By introducing big-M constraints, the bilinear term can be linearized, which can enable us to transform the sub-problem into an MISOCP. When $\mathbf{d}$ fluctuates within the interval $[\underline{\mathbf{d}}, \overline{\mathbf{d}}]$, introducing an auxiliary binary vector $\boldsymbol{\zeta}$ that satisfies $d_i = \underline{d}_i + \zeta_i \Delta d_i$ element-wise and an auxiliary variable $\gamma_i = \zeta_i \tau_i$, the above bilinear program is equivalent to the following MISOCP:

$$\max_{\mathbf{d} \in \mathcal{D}, \boldsymbol{\tau}, \underline{\boldsymbol{\lambda}}, \overline{\boldsymbol{\lambda}}, \boldsymbol{\xi}, \boldsymbol{\sigma}_l, \mu_l, \boldsymbol{\zeta}, \boldsymbol{\gamma}} -(\mathbf{A}\mathbf{x}^* + \underline{\mathbf{d}})^T \boldsymbol{\tau} + \mathbf{y}^T \underline{\boldsymbol{\lambda}} - \overline{\mathbf{y}}^T \overline{\boldsymbol{\lambda}} + (\mathbf{e} - \mathbf{C}\mathbf{x}^*)^T \boldsymbol{\xi} - \Delta \mathbf{d}^T \boldsymbol{\gamma} \quad (70)$$

$$\text{s.t.} \quad \boldsymbol{\tau} - M(\mathbf{1} - \boldsymbol{\zeta}) \le \boldsymbol{\gamma} \le \boldsymbol{\tau} + M(\mathbf{1} - \boldsymbol{\zeta}) \quad (71)$$

$$-M\boldsymbol{\zeta} \le \boldsymbol{\gamma} \le M\boldsymbol{\zeta} \quad (72)$$

$$\zeta_i \in \{0,1\} \quad (73)$$

$$\text{and } (67)\text{-}(69)$$

where $\mathbf{1}$ is a column vector whose elements are all equal to 1, $M$ is a very large positive number (e.g., $10^6$) and (71)-(72) are big-M constraints.

## C. C&CG Algorithm

The whole procedure to solve the two-stage robust reactive power optimization problem is summarized in Table I.

TABLE I  A SUMMARY OF THE C&CG ALGORITHM FOR ROBUST VOLT-VAR OPTIMIZATION PROBLEM

1. **Initialization** $LB := 0$, $UB := \infty$, $t := 0$. Set tolerance level $\varepsilon > 0$.
2. Solve the master problem to get an optimal dispatch scheme $\mathbf{x}^*$ and $\eta^*$. Then update $LB := \max\{LB, \eta^*\}$.

   **While** $UB - LB > \varepsilon$
   3. Given $\mathbf{x}^*$, solve the sub-problem to obtain an optimal objective $\eta(\mathbf{x}^*)$ corresponding to the worst-case scenario $\mathbf{d}^*$. Update $UB := \min\{UB, \eta(\mathbf{x}^*)\}$.
   4. Update $\mathbf{d}(t+1) := \mathbf{d}^*$. Create variables $\mathbf{y}(t+1)$ and add corresponding constraints (37)-(40) to the reduced master problem.
   5. Update $t := t+1$. Return to Step 1.

To be more readable, a schematic flowchart of the proposed algorithm is depicted in Figure 4. The interaction between the master problem and the sub-problem is annotated over the arrow. The master problem solves a reduced problem considering typical worst-case scenario set and then transfers the optimal dispatch $\mathbf{x}^*$ for slow control devices to sub-problem. The sub-problem simply solves the dual problem of optimal power flow, searches for the worst-case scenarios given the slow dispatch

$x^*$, and renew the typical worst-case scenario set in the master problem. The master problem updates the lower bound of the objectives, while the sub-problem updates the upper bound. The whole procedure terminates while the difference of these two bounds are sufficiently small.

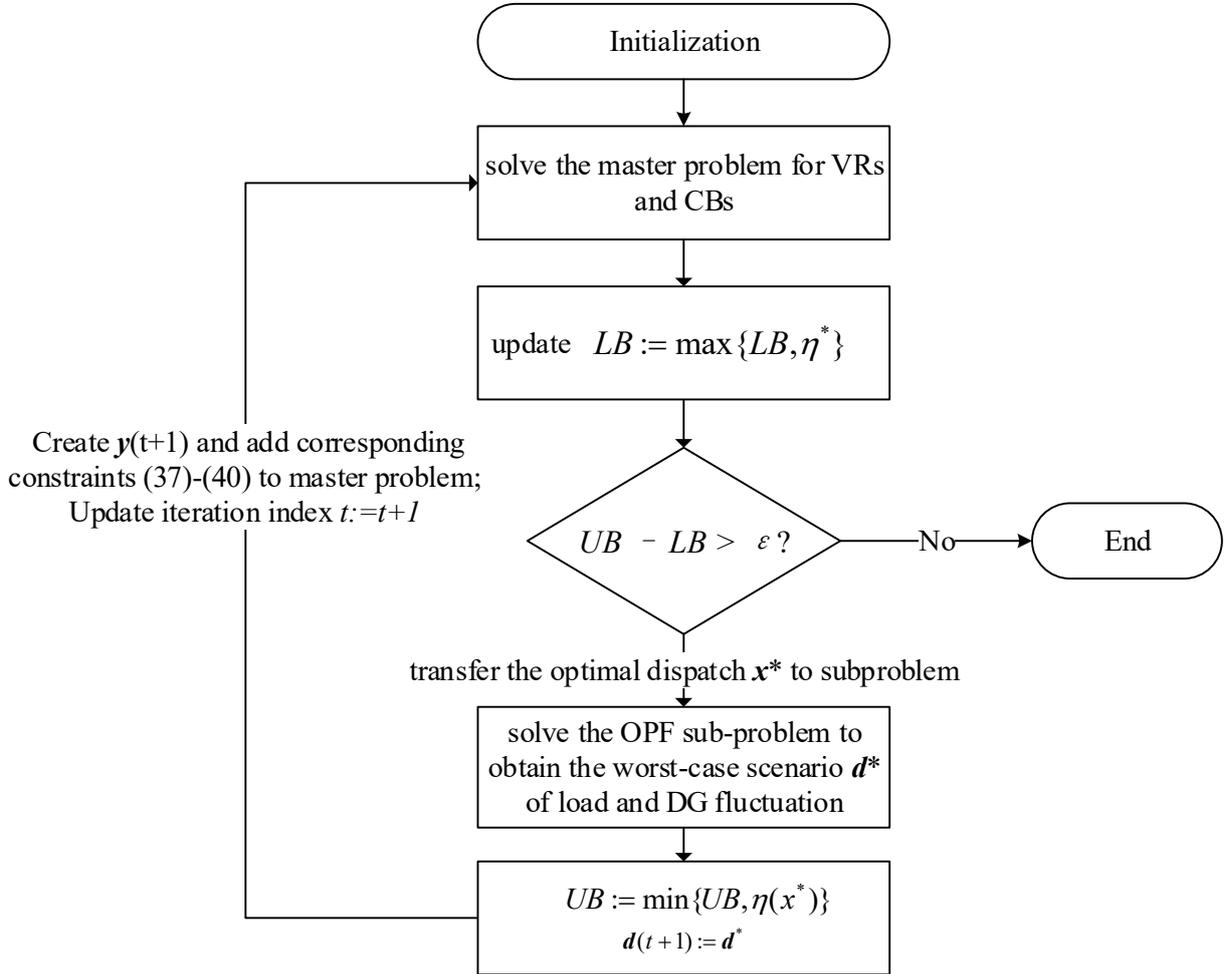

Figure 4 Flowchart of the proposed algorithm.

IV. SIMULATION RESULTS

In this section, we outline results from numerical experiments on a real feeder from Shandong distribution system in China [6][16], an IEEE 69-bus and a three-phase unbalanced IEEE 123-bus distribution systems [17]. These were conducted to enable us to evaluate the performance of the proposed robust model.

TABLE II CONFIGURATIONS OF TEST SYSTEMS

| System | Shandong 14-bus test feeder | IEEE 69-bus | IEEE 123-bus |
|---|---|---|---|
| DG | buses {5, 6, 7, 8, 9, 10, 11, 12, 13} | buses {19, 20, 26, 22, 54, 69, 34, 38} | buses {7, 12, 33, 47, 115, 98, 84, 19, 43, 61} |
| SVC | bus 14 | - | buses {94, 68} |
| VR | line 6 | line 10 | line 58 |

DG, distributed generations; SVC, static VAR compensators; VR, voltage regulator; MSC, capacitor bank/ mechanically switched capacitor

In the IEEE 69-bus system, there are eight DGs, one VR and five MSCs. In the IEEE 123-bus system, there are ten DGs, two SVCs, one VR and five MSCs. DG at bus 68 is operated at phase B, while DGs at buses at 61 and 94 are operated at phase A and phase B. Other equipment is operated in three phase. Details of the configuration is listed in Table II.

The admittance of the MSCs was set as 1.0 p.u.. The capacity of the SVCs was 0.3 p.u. The capacity of the DGs was 0.3 p.u. for both active and reactive power generation. Suppose that the DGs operate in maximum power point tracking (MPPT) mode, which attempts to generate the maximum allowable active power, and the reactive power is optimized [18]. The allowable turn ratio of the VR lie in the interval [0.95, 1.05] with step length 0.01. The operational limits for voltage magnitude are 0.9 p.u. and 1.042 p.u. [21].

The algorithms, namely the Deterministic Volt-VAR Optimization Model (DVO) and the two stage Robust Volt-VAR Optimization Model (RVO), were implemented using the Matlab program and solved with IBM ILOG CPLEX software (ver.12.5) [19]. The test environment was a laptop with an Intel Core i5-3210M 2.50-GHz processor and 8 GB of RAM.

The following uncertainty set is considered:

$$\mathcal{D} = \left\{ \begin{pmatrix} P_D \\ Q_D \\ P_G \end{pmatrix} \middle| \begin{array}{l} (1-v)P_D^{Base} \leq P_D \leq (1+v)P_D^{Base} \\ (1-v)Q_D^{Base} \leq Q_D \leq (1+v)Q_D^{Base} \\ (1-v)P_G^{Base} \leq P_G \leq (1+v)P_G^{Base} \end{array} \right\} \quad (74)$$

where the base loads $P_D^{Base}, Q_D^{Base}$ were taken from the data in the standard test archive and the volatility is denoted as $v$.

The simulation results are organized as follows. In section A, we first demonstrate the practical applicability issue of RVO. We use a real feeder in China and a scenario along with the base case to show the technical issues arising from DVO and how RVO can tackle with this issue. Then, both DVO and RVO were tested on a large number of scenarios generated by Monte Carlo simulation in section B, using two IEEE standard test archives. The results show that RVO can handle uncertainties in a single-time period. In section C, numerical tests were further performed in multiple-time period, and RVO can successively keep the voltage profiles within security region.

## A. Practical applicability of the proposed algorithm

The major drawback of DVO is firstly demonstrated through a real feeder from Shandong distribution system in China. Detailed data are all online available [16]. In this section, the base loads $P_D^{Base}, Q_D^{Base}$ were taken from the data in [16] and the volatility $v$ is set as 50%. In the base case, although the voltage magnitudes of all buses did not violate upper bound, voltage violation occurred in buses {3, 5-14} after fluctuation of loads and DG outputs (50% loads and 150% DG active power outputs in comparison to the base case), as shown in Figure 5. And this practical issue can be addressed by RVO, which can keep voltage security even after loads and DG outputs fluctuation as shown in Figure 6.

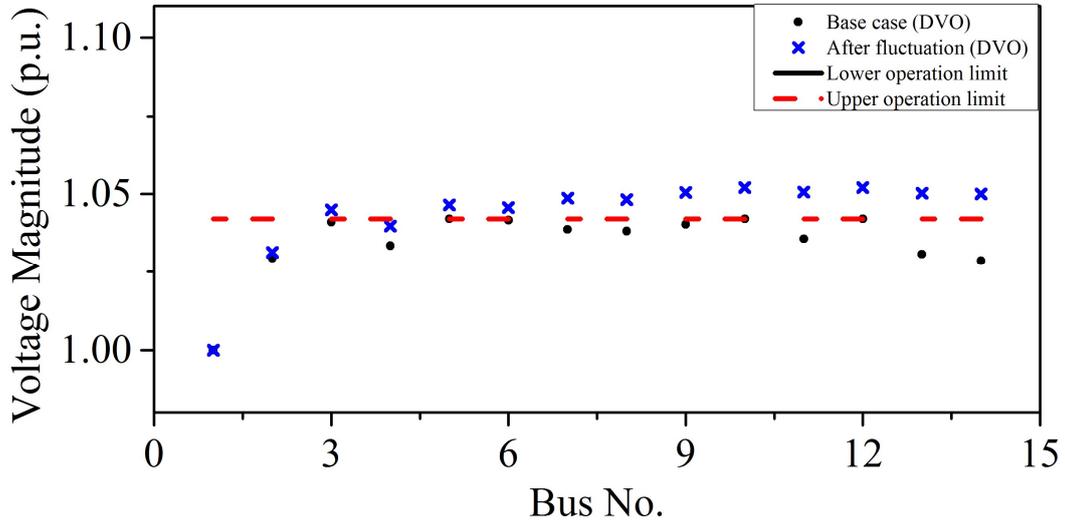

Figure 5  Simulated voltage magnitude of the convex deterministic Volt-VAR Optimization Model (DVO) in the Shandong Test Feeder.

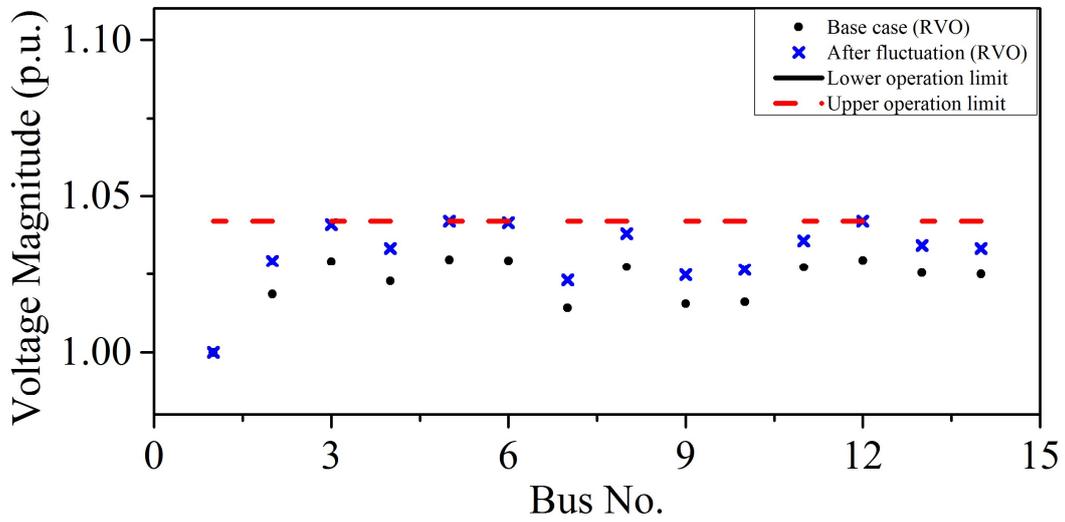

Figure 6  Simulated voltage magnitude of the two stage robust Volt-VAR Optimization Model (RVO) in the Shandong Test Feeder.

TABLE III    CONTROL STRATEGIES FOR SLOW DEVICES GENERATED BY RVO AND DVO IN SHANDONG TEST FEEDER (1: SWITCH ON, 0: SWITCH OFF)

|  | Devices | Strategy | |
|---|---|---|---|
|  |  | DVO | RVO |
| Dispatch schedule | MSC11 | 1 | 0 |
|  | MSC12 | 0 | 0 |
|  | MSC13 | 1 | 0 |
|  | MSC14 | 1 | 1 |
|  | VR6 | 1.02 | 0.99 |

Here, the reason why RVO differs from DVO is briefly discussed using the dispatch schedules generated by them in Table III. Since DVO does not consider any uncertainties, its objective is to minimize network losses by increasing the voltages profile for the base-case scenario, which would probably encounter voltage violation after fluctuation of DGs. RVO is comparatively conservative when making control decisions for slow control devices by considering the worst-case scenario, which make the subsequent fast control has the capability to eliminate the voltage violations.

## B. Comparison of handling uncertainties in a single-time period

In this section, the base loads $P_D^{Base}, Q_D^{Base}$ were taken from the data in [17] and the volatility $v$ is set to 10%. In total, 100 Monte Carlo runs were conducted independently to generate scenarios.

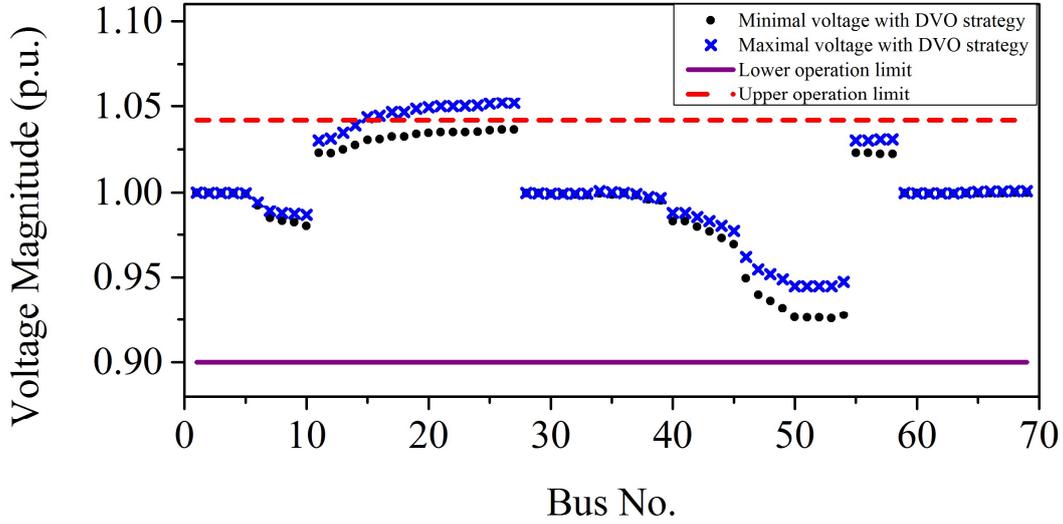

Figure 7 Simulated voltage magnitude of the convex deterministic Volt-VAR Optimization Model (DVO) in the IEEE 69-bus system.

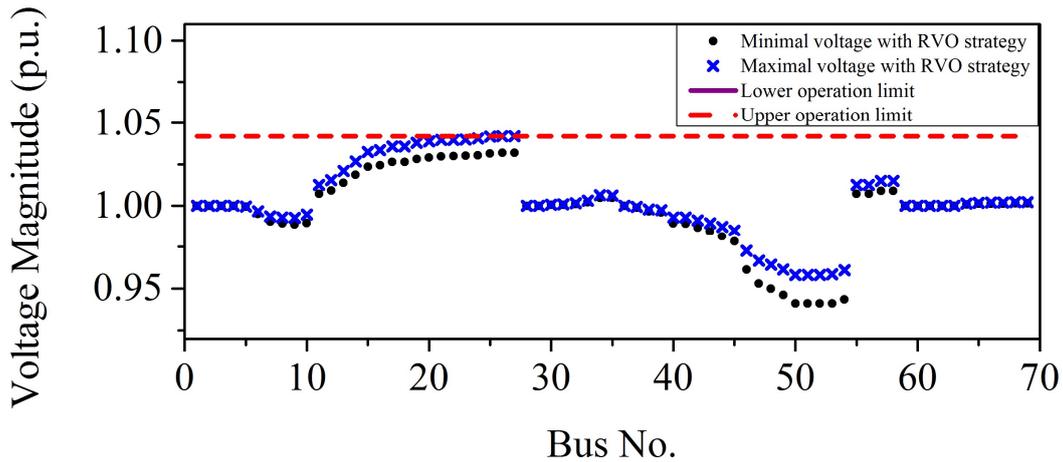

Figure 8 Simulated voltage magnitude of the two stage robust Volt-VAR Optimization Model (RVO) in the IEEE 69-bus system.

In simulation, the DVO made optimal dispatches for VRs and MSCs, and achieved optimal reactive power flow for forecasted loads and DG outputs. However, the dispatch could not accommodate fluctuating loads and DG outputs, and the simulated voltage magnitude varied between the blue "×" and the black circle in Figure 7. The slow control in the RVO considered all possibilities in the uncertainty set, and taken account each possible loads and active output of DGs. Therefore, the fast control enhanced optimality and feasibility. As a result, no voltage violation was observed in RVO simulations, which were prepared for the worst-case scenario, as shown in Figure 8.

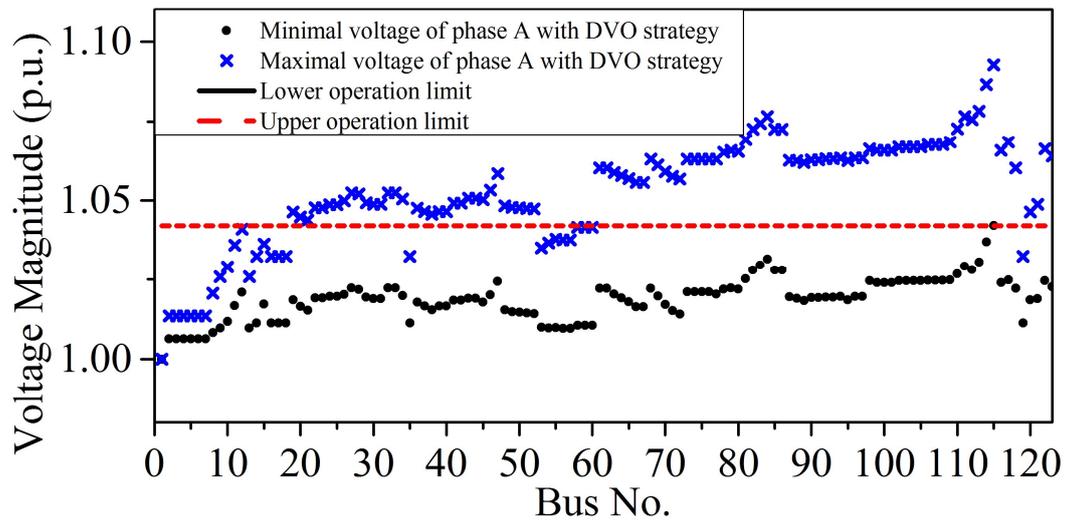

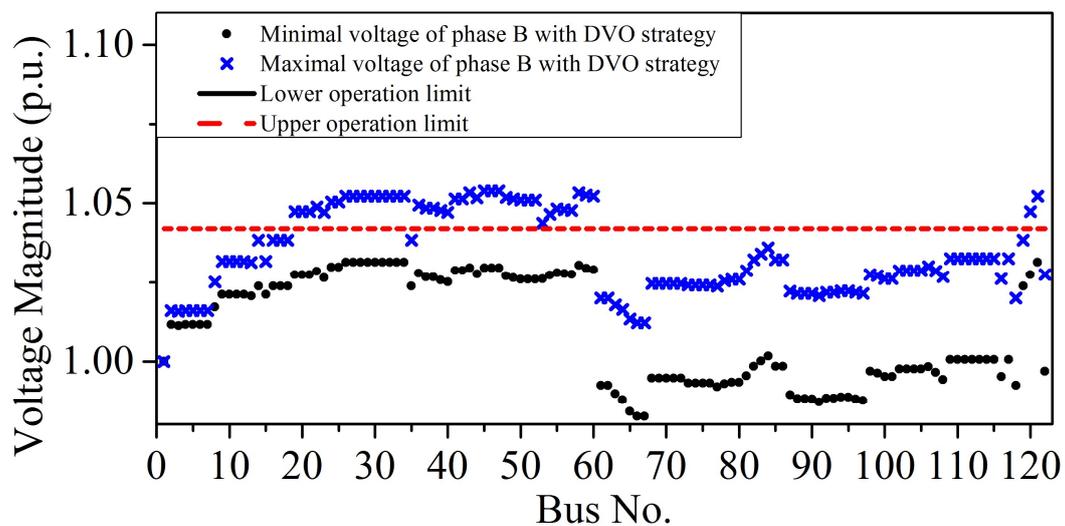

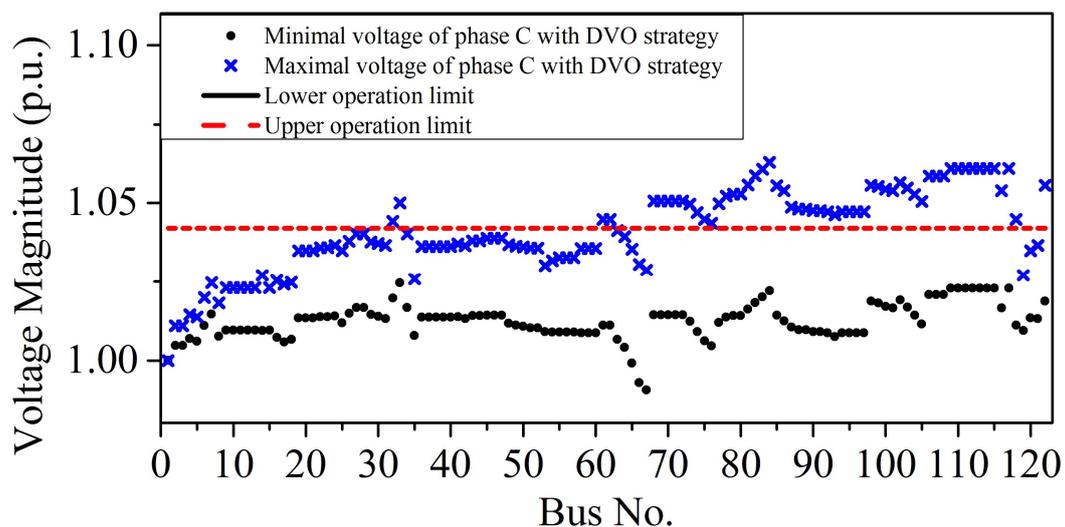

Figure 9 Simulated voltage magnitude of DVO in the IEEE 123-bus system for (a) phase A; (b) phase B; and (c) phase C.

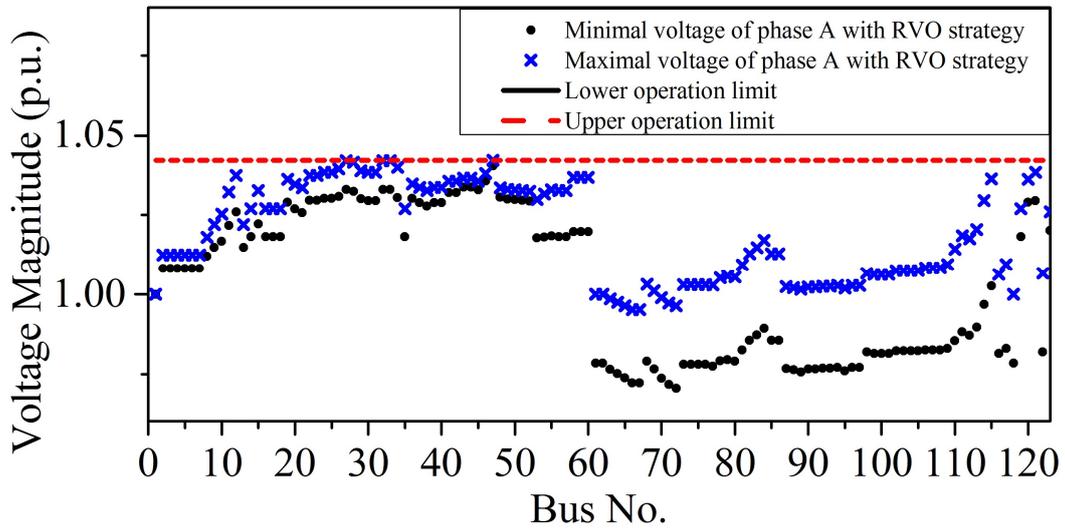

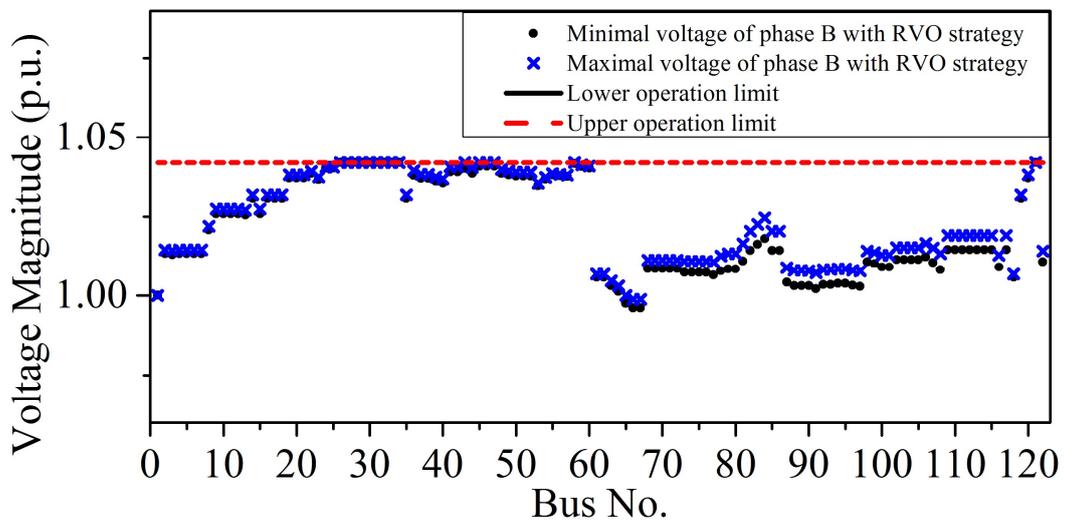

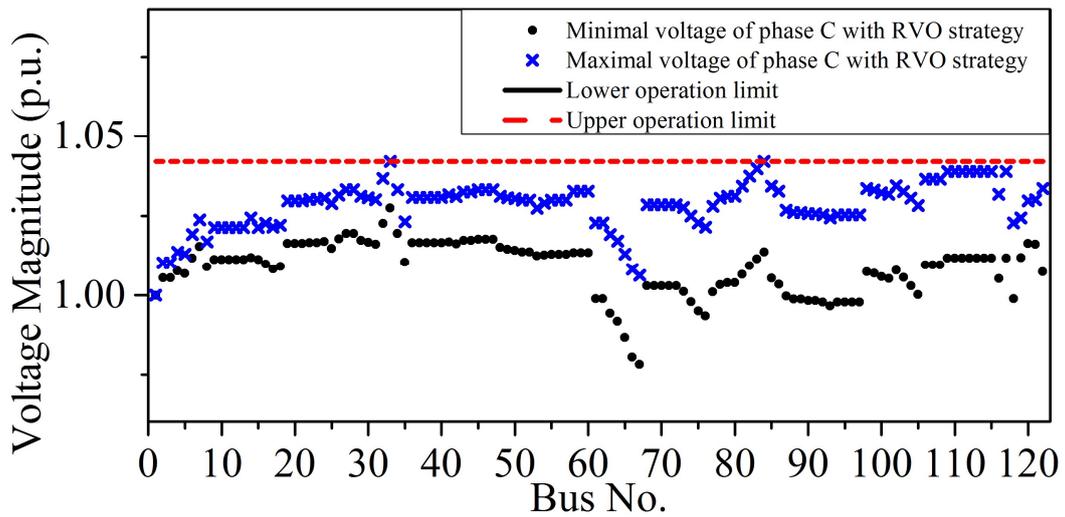

Figure 10 Simulated voltage magnitude of RVO in the IEEE 123-bus system for (a) phase A; (b) phase B; and (c) phase C.

In Figure 9, we see that the results for the IEEE 123-bus system were similar to those for the 69-bus system, except that voltage violation was more severe in this system with more DGs. Voltage violation was again eliminated by the RVO, as shown in Figure 10.

TABLE IV COMPARISON OF OVERALL PERFORMANCE BETWEEN DVO AND RVO

| System | DVO | | RVO | |
|---|---|---|---|---|
| | E[$P_{loss}$] (kW) | Failure Rate | E[$P_{loss}$] (kW) | Failure Rate |
| 69-bus | 81.9 | 44% | 92.2 | 0 |
| 123-bus | 212.0 | 68% | 213.0 | 0 |

DVO, convex deterministic Volt-VAR Optimization model; RVO, Two-Stage Robust Volt-VAR Optimization Model

The overall performance, including the expectation of the network loss and failure rate (i.e. the possibility of voltage violation), is summarized in Table IV. Note that there are a number of infeasible cases in the DVO simulations. To enable direct comparisons, only expected losses for feasible cases are considered in DVO. This means that the infeasible cases have been removed prior to calculating the expected losses for DVO. RVO can prevent voltage violation in uncertain conditions with a reasonable extra cost of optimality in comparison to DVO.

The uncertainty set (74) can be adjusted by changing the volatility $v$ of demand and DG output variation. Table V shows the sensitivity of the minimum worst-case power losses of the RVO with respect to the different volatility of the uncertainty set in the IEEE 123-bus system. It is observed that the worst-case power losses increases as the size of the uncertainty set gets larger. The result is intuitive. A larger uncertainty set will make things worse and incur a higher power loss. The iterations required to converge are 2~3 steps, which shows that RVO is numerically stable.

TABLE V COMPARISON OF RVO WITH DIFFERENT VOLATILITY OF THE UNCERTAINTY SETS IN IEEE 123-BUS SYSTEM

| Volatility $v$ | 0.02 | 0.04 | 0.06 | 0.08 | 0.1 |
|---|---|---|---|---|---|
| E[$P_{loss}$] (kW) | 208.9 | 210.1 | 211.2 | 212.0 | 213.0 |
| Iteration | 2 | 2 | 2 | 3 | 3 |

C. *Comparison of handling uncertainties in multi-time period*

The DVO and RVO were tested in multi-time period scenarios using the IEEE 123-bus System. Here, the DGs were assumed to be PV generators and their outputs were simulated using Hybrid optimization model for renewable energy (HOMER) software [20]. The standardized curves for the load and DG output is shown in Figure 11. The 100% load level is the base load available in the test archive in [17], while the 100% PV output is the maximal capacity.

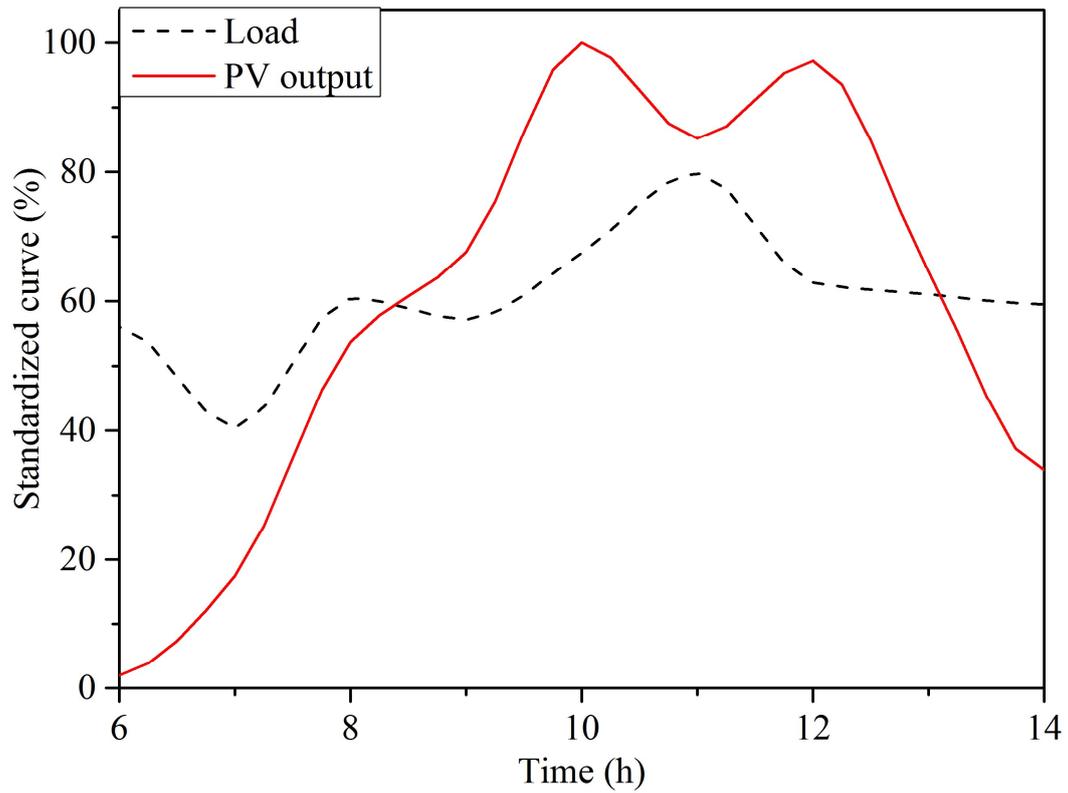

Figure 11    The standardized curves of the load and DG output

In the simulations, the DVO and RVO were compared using the load level and PV output level between 6:00 and 14:00. DVO made here-and-now decision for VRs, MSCs and optimal reactive power flow for current loads and DG outputs. When performing each slow control, an uncertainty set was constructed based on the extreme values of the injected active power over the whole test period. For each load and PV output, both algorithms adopted the fast controls. In both algorithms, the slow control interval was 2 hours and fast control interval was 15 minutes.

TABLE VI  CONTROL STRATEGIES FOR SLOW DEVICES GENERATED BY RVO AND DVO IN IEEE 123-BUS SYSTEM (1: SWITCH ON, 0: SWITCH OFF)

| Strategy | Devices | Dispatch schedule | | | |
|---|---|---|---|---|---|
| | | 6:00 | 8:00 | 10:00 | 12:00 |
| DVO | MSC4 | 1 | 0 | 0 | 1 |
| | MSC12 | 1 | 0 | 0 | 0 |
| | MSC19 | 1 | 0 | 0 | 0 |
| | MSC55 | 0 | 0 | 0 | 0 |
| | MSC93 | 0 | 0 | 0 | 0 |
| | MSC68 | 0 | 0 | 0 | 0 |
| | MSC43 | 1 | 1 | 1 | 1 |
| | MSC5 | 1 | 0 | 1 | 0 |
| | MSC119 | 0 | 0 | 0 | 0 |
| | VR58 | 1.01 | 1.01 | 1.03 | 1.03 |
| RVO | MSC4 | 0 | 1 | 0 | 0 |
| | MSC12 | 1 | 1 | 0 | 1 |
| | MSC19 | 1 | 1 | 0 | 1 |
| | MSC55 | 0 | 0 | 0 | 0 |
| | MSC93 | 0 | 0 | 0 | 0 |
| | MSC68 | 0 | 0 | 0 | 0 |
| | MSC43 | 0 | 1 | 1 | 1 |
| | MSC5 | 0 | 0 | 0 | 0 |
| | MSC119 | 0 | 0 | 0 | 0 |
| | VR58 | 0.96 | 1.03 | 1.04 | 1.00 |

A major reason for overvoltage is reverse power flow [21]. This problem is likely to occur in DVO when the growth of the PV outputs exceeds the growth of loads. This was especially the case in the period from 6:00 to 10:00 and from 11:00 to 12:00 in Figure 11. Table VI summarized the dispatch results for slow control using the DVO and RVO strategies for the IEEE 123-bus system. Figure 12 depicts the number of overvoltage buses in the 123-bus system. Even with sequential fast controls, some buses operating beyond the upper voltage limits in the cases of DVO. However, in the cases of RVO, no voltage violations were observed for the entire time period investigated (6:00-14:00).

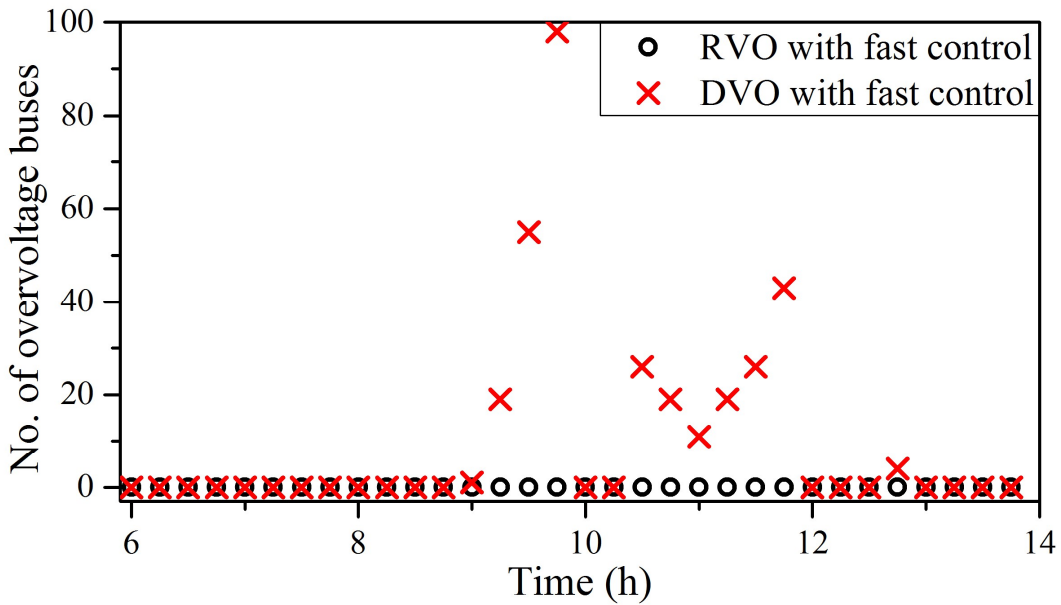

Figure 12    The evolution of overvoltage buses in the IEEE 123-bus system.

V. CONCLUSION

We described a dual time-scale coordinated robust reactive power and voltage optimization model. This model can handle the uncertainties in predicted load and distributed generation. The nonconvex optimization problem is first convexified using second-order conic relaxation for branch flow, piecewise linearization for VRs and an equivalent transformation for MSCs. Then a C&CG algorithm was used to solve the convexified robust model. The simulation results show that the solution of the deterministic model is not always feasible and voltage violations may occur. The robust model can ensure security with regard to all possible scenarios limited by the uncertainty sets with little compromise in terms of losses. Due to the MADPT constraints, the dispatch of the slow controlling devices should be formulated as a dynamic optimization problem. To be applicable, this problem is always simplified into a sequence of independent static optimization problems. This means that we need to schedule the operating times of the slow controlling devices in advance. Future research should address the question of exactly how to schedule these operating time. Development and implementation of a software that can be directly deployed in the operation center is also an important future work.

VI. ACKNOWLEDGMENTS

This work was supported in part by the National Key Basic Research Program of China (Grant. 2013CB228205), in part by the National Science Foundation of China (Grant. 51477083)

REFERENCES

[1]   L. Franchi, M. Innorta, P. Marannino, and C. Sabelli, "Evaluation of economy and/or security oriented objective functions for reactive power scheduling in large scale systems," *Power Apparatus and Systems, IEEE Transactions on,* vol. PAS-102, no. 10, pp. 3481-3488, 1983.
[2]   C. D'Adamo, S. Jupe, and C. Abbey, "Global survey on planning and operation of active distribution networks - Update of CIGRE C6.11 working group activities," in *Electricity Distribution - Part 1, 2009. CIRED 2009. 20th International Conference and Exhibition on*, 2009, pp. 1-4.
[3]   P. M. De Oliveira-De Jesus, E. D. Castronuovo, and M. T. Ponce de Leao, "Reactive power response of wind generators under an incremental network-loss allocation approach," *Energy Conversion, IEEE Transactions on,* vol. 23, no. 2, pp. 612-621, 2008.
[4]   H. Ahmadi, J. R. Marti, and H. W. Dommel, "A framework for Volt-Var optimization in distribution systems," *Smart Grid, IEEE Transactions on,* vol. 6, no. 3, pp. 1473-1483, 2015.
[5]   B. Zhang, A. Y. S. Lam, A. D. Dominguez-Garcia, and D. Tse, "An optimal and distributed method for voltage regulation in power distribution systems," *Power Systems, IEEE Transactions on,* vol. 30, no. 4, pp. 1714-1726, 2015.
[6]   W. Zheng, W. Wu, B. Zhang, H. Sun, and Y. Liu, "A fully distributed reactive power optimization and control method for active distribution networks," *Smart Grid, IEEE Transactions on,* vol. PP, no. 99, pp. 1-1, 2015.
[7]   A. T. Saric, Stankovic, x, and A. M., "A robust algorithm for Volt/Var control," in *Power Systems Conference and Exposition, 2009. PSCE '09. IEEE/PES*, 2009, pp. 1-8.
[8]   Y. Wang, W. Wu, B. Zhang, Z. Li, and W. Zheng, "Robust voltage control model for active distribution network considering PVs and loads uncertainties," in *Power & Energy Society General Meeting, 2015 IEEE*, 2015, pp. 1-5.
[9]   N. Daratha, B. Das, and J. Sharma, "Robust voltage regulation in unbalanced radial distribution system under uncertainty of distributed generation and loads," *International Journal of Electrical Power & Energy Systems,* vol. 73, pp. 516-527, 12// 2015.
[10]  W. Wu, B. Zhang, and K. L. Lo, "Capacitors dispatch for quasi minimum energy loss in distribution systems using a loop-analysis based method," *International Journal of Electrical Power & Energy Systems,* vol. 32, no. 6, pp. 543-550, 2010.
[11]  Z. Li, W. Wu, B. Zhang, and B.Wang, "Adjustable robust real-time power dispatch with large-scale wind power integration," *Sustainable Energy, IEEE Transactions on,* vol. 6, no. 2, pp. 357-368, 2015.
[12]  M. Farivar and S. H. Low, "Branch flow model: relaxations and convexification (Part I)," *Power Systems, IEEE Transactions on,* vol. 28, no. 3, pp. 2554-2564, 2013.
[13]  M. E. Baran and F. F. Wu, "Network reconfiguration in distribution systems for loss reduction and load balancing," *Power Delivery, IEEE Transactions on,* vol. 4, no. 2, pp. 1401-1407, 1989.
[14]  B. Liu, F. Liu, S. Mei, and Y. Chen, "Optimal reactive power flow with exact linearized transformer model in distribution power networks," in *Control and Decision Conference (CCDC), 2015 27th Chinese*, 2015, pp. 5562-5567.
[15]  B. Zeng and L. Zhao, "Solving two-stage robust optimization problems using a column-and-constraint generation method," *Operations Research Letters,* vol. 41, no. 5, pp. 457-461, 9// 2013.
[16]  Distribution system in Shandong province, China [Online].https://drive.google.com/file/d/0BzVXltCTbTuyZU1aY0MzVUZGTGc/view?usp=sharing
[17]  Distribution Test Feeders [Online]. Available: http://ewh.ieee.org/soc/pes/dsacom/testfeeders/
[18]  T. Esram and P. L. Chapman, "Comparison of photovoltaic array maximum power point tracking techniques," *Energy Conversion, IEEE Transactions on,* vol. 22, no. 2, pp. 439-449, 2007.
[19]  Cplex [Online]. Available: http://www-01.ibm.com/software/integration/optimization/cplexoptimizer/

[20] HOMER [Online]. Available: www.nrel.gov/docs/fy04osti/35406.pdf
[21] Y. Wang, P. Zhang, W. Li, W. Xiao, and A. Abdollahi, "Online overvoltage prevention control of photovoltaic generators in microgrids," *Smart Grid, IEEE Transactions on,* vol. 3, no. 4, pp. 2071-2078, 2012.